\documentclass[twocolumn,prl,preprintnumbers,amsmath,amssymb]{revtex4}

\usepackage{bm}
\usepackage{times}

\usepackage[dvips]{graphicx}
\usepackage[dvips]{color}
\newcommand{\grad}{\bm{\nabla}}
\newcommand{\vct}[1]{\mathbf{#1}}

\begin{document}
\title{Shear flow pumping in open microfluidic systems}

\author{Markus Rauscher}
\author{S. Dietrich}
\affiliation{Max-Planck-Institut f{\"u}r Metallforschung,
Heisenbergstr. 3, 70569 Stuttgart, Germany} \affiliation{Institut
f{\"u}r Theoretische und Angewandte Physik, Universit{\"a}t Stuttgart,
70569 Stuttgart, Germany}
\author{Joel Koplik}
\affiliation{Benjamin Levich Institute and Department of Physics,
City College of the City University of New York, New York, NY
10031, USA}

\date{\today}

\begin{abstract}
We propose to drive open microfluidic systems by shear in a covering fluid
layer, e.g., oil covering water-filled chemical channels. The
advantages as compared to other means of pumping are
simpler forcing and prevention of evaporation of volatile components. 
We calculate the expected throughput for straight channels
and show that devices can be built with off-the-shelf technology. 
Molecular dynamics simulations suggest that this concept is scalable
down to the nanoscale.
\end{abstract}

\maketitle


Progressive miniaturization and integration of chemical and physical
processes into microfluidic systems, so-called ``chemical chips'', is
expected to permit cheap mass production and to facilitate the handling
of much smaller quantities 
than standard
laboratory equipment \cite{giordano01,mitchell01,stone01}\/. Most currently
built microfluidic systems contain the liquids in closed pipes and use
body forces (e.g., gravity or centrifugal forces) \cite{gustafsson03}, electroosmosis,
capillary wicking, or simply applied pressure to generate flow.
Driving
these systems is relatively straightforward, but there are major drawbacks.
Small pipes get easily clogged (in particular while processing 
biological fluids which contain large molecules such as proteins or DNA),
driving gets increasingly difficult with reduced cross section,
and fabrication more challenging.

Open microfluidic systems are a second line of development which
can overcome
these problems. As demonstrated experimentally \cite{darhuber01,gau99} and
theoretically \cite{dietrich05,koplik06a}, fluids can be guided on flat
solid surfaces by wettability patterns, e.g., hydrophilic stripes on an
otherwise hydrophobic substrate. Several methods to generate flow in such
''chemical channels`` have been discussed: capillary forces (i.e., 
wicking into channels or motion due to wettability gradients)
\cite{darhuber01}, electrowetting 
\cite{srinvansan04,zeng04}, thermally generated Marangoni forces (e.g., in
arrays of local heaters) \cite{farahi04,kotz04},
body forces, surface acoustic waves
\cite{guttenberg05}, or combinations of these. While capillary forces can only
induce flow over relatively short distances, electrode and heater arrays
require complex sample structures.
Surface acoustic waves can only drive
drops which are large compared with the wavelength (in general on the micron
scale), and high rotation velocities are required to obtain
sufficiently strong centrifugal forces.

\begin{figure}
\includegraphics[width=\linewidth]{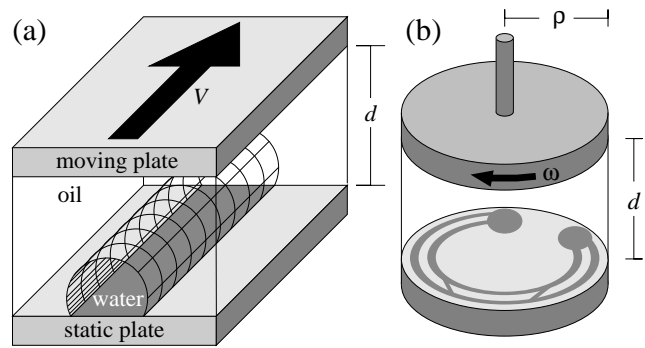}
\caption{\label{fig:setup} (a) Flow in a water filled chemical channel (dark
stripe on the static plate) can be induced by shear in the oil filling the
gap between the static bottom plate and the moving top cover. (b) 
In a device the motion of the top with respect to the bottom plate
can be realized by two circular discs rotating relative to each
other. The pattern on the bottom plate indicates two reservoirs
(circles) connected by two channels with two interconnections.  }
\end{figure}

As illustrated in Fig.~\ref{fig:setup} we propose to use shear,
generated in 
a covering fluid layer (e.g., oil on water),
to induce flow over long length scales, e.g., across
the whole device. A top plate
moving with respect to the bottom patterned substrate generates shear in the oil
which fills the gap between the two plates and, in turn, the oil will induce
flow in a chemical channel covered with, e.g., water.
This principle can be
realized in a device with two circular plates rotating relative to each
other and the chemical patterning can be implemented with standard printing
techniques \cite{abbott99,zhao02}\/. 
Design and scalability of shear driven systems is similar to body
force driven devices and the same type of surface tension driven
pearling instabilities occur in straight channels \cite{koplik06a}\/.
In both cases pearl formation increases throughput through the channels.
This driving principle can be
extended easily to water confined in open topographic channels
\cite{seemann05}, although it will be more efficient in open
chemical channel systems as there the water has less contact with
the substrate (due to the lack of sidewalls)\/.

In the following we calculate the throughput for a shear driven straight
chemical channel in two simple cases. Using molecular dynamics
simulations of a system of two immiscible model fluids we
demonstrate that shear driving can be downscaled to the nanometer
scale. We conclude by discussing how the principle of shear
driving can be realized in a device of the size of a miniature
computer hard disk (microdrive)\/.

In order to estimate the possible throughput
we consider the simple case of a straight and
homogeneously filled chemical channel aligned parallel to the shear flow in the
covering oil layer. We assume that the three phase contact line is pinned at
the channel edges. In microfluidic
devices the Reynolds number is typically small so that one has
Stokes flow. For two
situations
one can solve the corresponding hydrodynamic equations
analytically: (i) 
the water-oil interface meets the substrate surface at a
contact angle of $\theta = 90^\circ$,
(ii) the viscosities of the oil and water are equal.

\begin{figure}
\includegraphics[width=\linewidth]{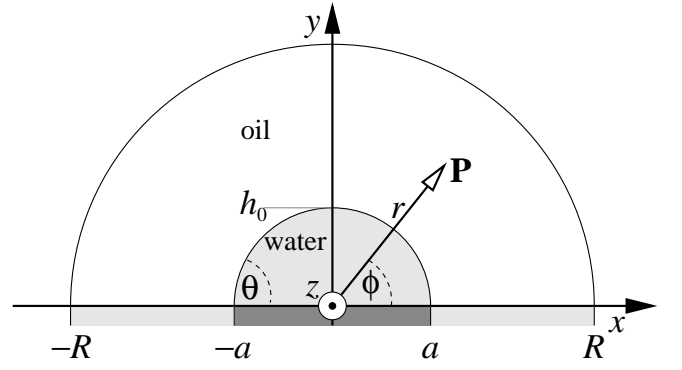}
\caption{\label{fig:drive} Cross section of a chemical
channel of width $2\,a$ filled up to a contact angle 
$\theta$ (here $\theta = 90^\circ$) and filling height $h_0$\/. 
In order to be able to use polar coordinates $\vct{P}=(x=r\,\cos\phi,\,
y=r\,\sin\phi)$ we fix the shear rate at
$r=R$ and consider the limit $R/a\to\infty$\/. 
}
\end{figure}

\begin{figure}
\includegraphics[width=\linewidth]{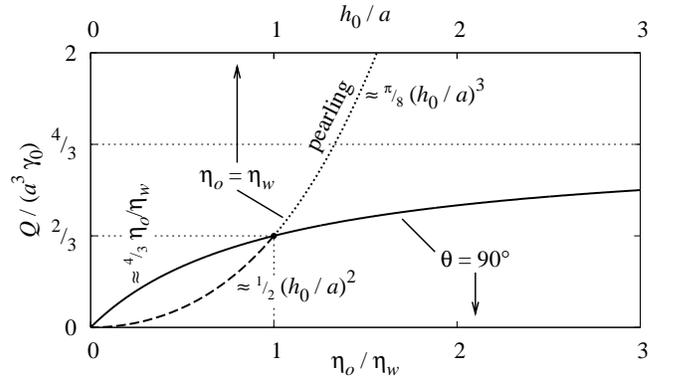}
\caption{\label{fig:flux} Normalized flux $Q$ for
$\theta=90^\circ$ as a function of $\eta_o/\eta_w$ (full line) and
for $\eta_o = \eta_w$ as a function of $h_0/a$ (dashed and dotted
line)\/. The asymptotics for small and large arguments are
indicated. For $h_0/a>1$ the liquid ridge is unstable with respect
to pearling (dashed line)\/.}
\end{figure}

Figure~\ref{fig:drive} shows a
cross section in a plane orthogonal to the chemical channel of width $2\,a$\/. 
Far from the channel
we assume a constant shear rate $\dot{\gamma}_0$\/. We seek
solutions of the Stokes equation with the flow parallel to the
channel (i.e., 
normal to the plane depicted in Fig.~\ref{fig:drive})\/. 
In this case, the pressure in the system is constant
and the velocity component $w$ parallel to the
channel satisfies a Laplace equation 
\begin{equation}
\label{eq:laplace}
\grad\cdot \eta_{w/o}\,\grad w = 0
\end{equation}
in both the water and the oil,
with viscosity $\eta_w$ and $\eta_o$, respectively\/. By
symmetry, the other velocity components are zero.
For Couette flow in a homogeneous medium between two parallel moving
plates (Fig.~\ref{fig:setup}(a)) the solution of
Eq.~(\ref{eq:laplace}) is $w=\dot{\gamma}_0\,y$, with a spatially
and temporally constant shear rate $\dot{\gamma}_0$. 
In order to be able to use polar
coordinates with the origin at the center of the chemical channel 
we take this solution as a boundary condition on an outer circle
with radius $R\gg a$, i.e., $w|_{r=R}=\dot{\gamma}_0\,R\,\sin\phi$
(Fig.~\ref{fig:drive})\/.
We apply a no-slip boundary condition at the liquid-substrate interface
for both fluids, i.e., $w|_{\phi=0,\pi}=0$\/. At the interface between the two
liquids 
the difference between the
normal components of the stress is given by the Laplace pressure.
For the
geometry considered here this reduces to the condition that the interface
has a constant curvature, i.e., it is a semicircle of radius $r=a$\/.
The velocity field, $w|_{r=a-0}= w|_{r=a+0}$, as well as 
the tangential stress components, $\eta_w\,\frac{\partial w}{\partial
r}|_{r=a-0}=\eta_o\,\frac{\partial w}{\partial r}|_{r=a+0}$, are continuous.
For $R/a \gg 1$, the corresponding solution of
Eq.~(\ref{eq:laplace}) is
\begin{equation}
w(r,\phi)=\dot{\gamma}_0\,\left\{
\begin{array}{cl}
\frac{2\eta_o}{\eta_o+\eta_w}\,r\,\sin\phi, & 
0<r<a\\
( r+\frac{\eta_o-\eta_w}{\eta_o+\eta_w}\,\frac{a^2}{r} )\,\sin\phi
,&
a<r
\end{array}
\right. .
\end{equation}
For $R/a\gg 1$ the flux (or throughput) $Q=\int_{A} w\,d^2r$, with
the channel cross section $A$, is given by
\begin{equation}
\label{eq:through}
Q_{90^\circ} = 
\frac{4}{3}\,a^3\,\frac{\eta_o\,\dot{\gamma}_0}{\eta_o+\eta_w}
= \frac{2}{3}\,a^2\,w^*_{90^\circ},
\end{equation}
see Fig.~\ref{fig:flux}\/. It increases with $\eta_o/\eta_w$\/.
The throughput can be characterized also by the apex velocity $w^*$, 
i.e., the fluid velocity at $\phi=0$ and $r=a$:
\begin{equation}
\label{eq:apex}
w^*_{90^\circ} = \frac{2\,a \eta_o\,\dot{\gamma}_0} {\eta_o+\eta_w}.
\end{equation}

For $\eta_o =\eta_w$ the flow field
is not influenced by the presence of the liquid-liquid interface as long
as the former is translationally invariant along the channel.
Accordingly, the velocity field is
\begin{equation}
w(y)=\dot{\gamma}_0\,y.
\label{eq:profile}
\end{equation}
Eq.~(\ref{eq:profile}) implies as water flux through a channel
with a cross section forming a circular segment with filling
height $h_0(\theta) = a\,(1-\cos\theta)/\sin\theta$
(see Fig.~\ref{fig:drive})
\begin{multline}
\label{eq:eqthrough}
Q_{\eta\eta} = \dot{\gamma}_0\,\Bigg\{ 
\frac{2}{3}\,a^3 + \frac{a\,\left(h_0^2-a^2\right)^2}{4\,h_0^2}
\\+ \frac{\left(h_0^4-a^4\right)\,\left(h_0^2+a^2\right)}{8\,h_0^3}\,
\left[ \frac{\pi}{2}+\arcsin \frac{h_0^2-a^2}{h_0^2+a^2}
\right]\Bigg\}
\end{multline}
so that $Q_{\eta\eta}(h_0=a) = Q_{90^\circ}(\eta_o=\eta_w)$ (see
Fig.~\ref{fig:flux})\/. 


In order to suggest that shear driven microfluidics can be miniaturized
to the nano-scale, we have performed molecular dynamics simulations 
as a well suited approach in this regime.
To a large extent we have used standard techniques 
\cite{allen87,frenkel02,koplik95} 
with the implementation of the dynamics on 
wetting stripes following Ref.~\cite{koplik06a}\/.
The model system consists of two immiscible monatomic 
fluids and two types of wall atoms, 
with Lennard-Jones interactions $V_{ij}(r)=4\epsilon
\left[(r/\sigma)^{-12} -A_{ij} (r/\sigma)^{-6}\right]$ truncated
at $r=2.5\,\sigma$, where $r$ is the
interatomic separation, $\sigma$ measures the size of the repulsive
core, $\epsilon$ is the strength of the potential,
and $A_{ij}=A_{ji}$ is a dimensionless parameter that controls the
attraction between atoms of the four atomic species $i$ and $j$\/.  
The temperature is set to $T=1.0\,\epsilon/k_B$ by a Nos\'e-Hoover
thermostat. The liquids
are confined between solid walls at a distance $50.7\,\sigma$,
each composed of a layer of fcc unit cells (with lattice constant
$1.55\,\sigma$)
whose atoms are tethered to lattice sites with a stiff linear spring.
Periodic boundary conditions are applied in the two lateral directions.
Couette flow is achieved by translating the tether sites of the
upper wall at a constant velocity. Most of the interaction
strengths $A_{ij}$ have the standard value $1.0$, but $A_{ij}=0$
if $i$ and $j$ refer to different liquids, or the inner liquid
(the "water") and the non-wetting part of the wall, or the outer
liquid (the "oil") and the wetting part of the wall, or to wall
atoms. With these choices, the two liquids are completely
immiscible. Because the interactions among atoms of each of the two
species are identical, they have equal density and viscosity. A
number of different system sizes have been simulated, all giving
similar results;  the largest contained about 1.25 million atoms
in total, with wetting stripe dimensions $16.7\sigma\times 547\sigma$\/.  

\begin{figure}
\begin{center}
\includegraphics[width=\linewidth]{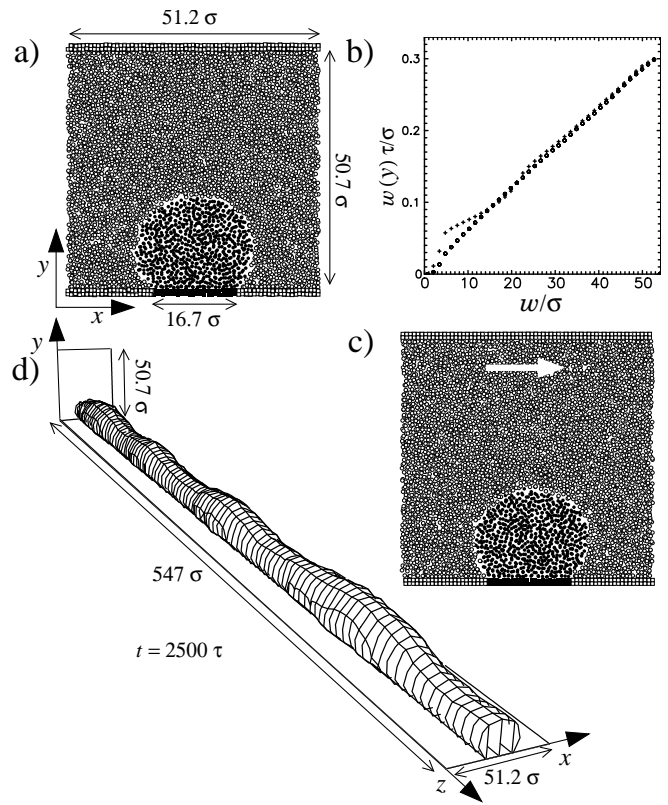}
\caption{\label{fig:md} (a) 
Cross sectional view
of the molecular configuration at early times before flow commences.
The inner stripe of substrate atoms (full boxes) attracts the
inner liquid atoms (full circles), while the outer substrate atoms (open
boxes) attract the outer liquid atoms (open circles)\/. Atoms
located within a slab of thickness $2.5\,\sigma$ are shown.
(b) Velocity profiles of the shear flow at the side of the simulation
box (o) and at the center of the channel (+), in slabs cutting through the
outer liquid and through the liquid ridge, respectively.
The upper solid wall is moved in $z$-direction  
at a velocity $0.3\,\sigma/\tau$.
(c) Shear at an oblique angle with a component orthogonal to the wetting stripe results in a tilted shape, but
the inner liquid ("water") is still transported along the channel.
(d) Chemical channels filled to more than $\theta = 90^\circ$ are
also unstable with respect to pearl formation under shear
(snapshot of the liquid-liquid interface at $t=2500\,\tau$)\/.} 
\end{center}
\end{figure}

After a short equilibration period 
the inner liquid occupies a roughly cylindrical region
pinned on one side to the wetting stripe (see
Fig.~\ref{fig:md}(a))\/. 
We note that the liquid-liquid interface is much smoother
than the previously studied liquid-vapor interfaces above wetting
stripes
\cite{koplik06a}\/. If the upper solid wall is moved
parallel to the stripe, a two-fluid Couette flow results.
Velocity profiles in two different vertical slabs parallel to the
$yz$-plane are shown in Fig.~\ref{fig:md}(b)\/. The velocities are
averaged over the slab width of $4\,\sigma$ and the full system 
length, and a time
interval of $250\,\tau$, with $\tau=\sigma\,\sqrt{m/\epsilon}$ and
the atom mass $m$\/. On the sides of the simulation box, where
only the outer liquid and the non-wetting solid are present, the
profile (o) varies linearly aside from a weak bending near the 
walls. In the center of the simulation box the velocity (+) shows
some additional deviation from linearity in the inner liquid
region. There is liquid-liquid interfacial slip (as expected for
the strongly immiscible liquids used here \cite{koplik06b}) and
some pinning behavior at the wetting stripe.  In both cases, the
velocity near both the stationary and moving wall exhibits a
no-slip behavior, although extrapolations of the velocity profile
from the interior of the channel would give rise to an apparent slip.

Under these circumstances
the anchoring of the inner liquid at the wetting stripe is very
stable, and we have observed 
that the inner liquid remains atop the stripe even if the
forcing occurs at oblique angles. As an extreme example, if the top plate
moves in the $x$-direction with the same velocity as above, the snapshot of
the atomic positions in Fig.~\ref{fig:md}(c) shows the inner fluid
to be only modestly displaced to the side. However, in general,
the interfacial shape is not stable, because the
configuration is subject to a surface tension driven instability as long as 
the channel is long enough and filled to more than a $90^\circ$
contact angle \cite{koplik06a}. In that case, a periodic surface
undulation develops, as shown in Fig.~\ref{fig:md}(d), which
eventually leads to propagating liquid pearls.


In order to discuss the applicability of our concept (see
Fig.~\ref{fig:setup}), we estimate the
throughput in a potential device. One prominent application is 
handling biological fluids which contain DNA fragments with a radius of
gyration of about $100$~nm\/. We therefore consider a channel 
width $2\,a = 1\,\mu\text{m}$ and a plate spacing $d=0.1$~mm\/. 
These are dimensions already
considered for chemical channels in slit pores \cite{lam02}\/. 
Modern miniature hard disks (so-called
Microdrives$^{\bigcirc\hspace{-1.35ex}\text{\sc r}}$ with a storage capacity
of up to 8~GB) have a disk radius $\rho$ (see
Fig.~\ref{fig:setup}(b)) of about one centimeter and rotate at
$\nu=3600\,\text{rpm}=60\,\text{Hz}$
(see, e.g., Hitachi Global Storage Technologies
{http://www.hgst.com})\/. 
The whole device has a size of $42.8\,\text{mm}
\times 36.4\,\text{mm} \times 5.0 \,\text{mm}$ so that building on this
technology might lead to microfluidic devices of similar dimensions.
For $\rho=1\,\text{cm}$ and $\omega=2\,\pi\,\nu=400\,\text{Hz}$ the
relative velocity of the discs at their perimeter is $v=\rho\,\omega =
4\,{\text{m}}/{\text{s}}$ and the shear rate is $\dot{\gamma}_0 = v/d = 
4\times 10^{4}\,{\text{s}}^{-1}$\/. Assuming $90^\circ$ contact
angles at the channel edges and that the viscosities of the two
fluids are equal Eq.~(\ref{eq:apex}) renders an apex velocity
$w^*_{90^\circ} = a\,\dot{\gamma}_0 = 2\,{\text{cm}}/{\text{s}}$\/.

In an inertia-driven situation the velocity at the surface of a
planar film of thickness $a$ driven parallel to the substrate
surface can be used as an order of magnitude 
estimate for the apex velocity  $w_g^* =
{a^2\varrho_w\,g}/(2\,\eta_w)$ with the mass density 
$\varrho_w$ of the inner fluid and the acceleration $g$\/. Thus for
$\eta_w=10^{-3}\,{\text{kg}}/({\text{m}\,\text{s}})$ and
$\varrho_w=10^3\,\text{kg}/\text{m}^3$ an apex velocity
$w_g^*=2\,{\text{cm}}/{\text{s}}$ requires an acceleration of
$g=1.6\times 10^5\,{\text{m}}\,{\text{s}^{-2}}$\/. Such accelerations
$\rho\,\omega^2$ can be
achieved with centrifugal forces at a radius of $1\,\text{cm}$ at 
$3.8\times 10^4~\text{rpm}$ 
as compared with a spin coater with
$10^4~\text{rpm}$, a turbo pump with $6\times 10^4~\text{rpm}$, and a 
dentist's drill with $3\times 10^5~\text{rpm}$\/. 

In order to assure laminar flow, the Reynolds number has to be
small, both in the channel and in the oil. With the values for
water one has $\text{Re}_w = {\varrho_w\,w^*\,a}/{\eta_w} =
10^{-2}$ in the channel. The relevant length scale for the
oil is $d$ and the velocity scale is $\rho\,\omega$ so that for 
$\varrho_o\approx \varrho_w$ and $\eta_o\approx
10^3\,\eta_w$ one obtains a similarly low Reynolds number for the
outer liquid, leading to laminar flow.

Viscous heating of the oil is a possible concern. For the
parameters discussed above, the power input into the device is of
the order of $\eta_w\,\dot{\gamma}_0^2\,d\,\pi\,\rho^2/4\approx
13\,\text{mW}$ so that cooling of the device is not expected to
be a problem.


In summary, we propose to drive flow in open microfluidic systems by shear in a
covering liquid of equal or higher viscosity, e.g., water filled channels
covered with oil. As a positive side effect, the oil prevents evaporation. 
The analysis of the Stokes flow for channels
filled to a $90^\circ$ contact angle at the channel edges shows
(see Fig.~\ref{fig:flux}) how
increasing the viscosity of the covering liquid leads to an
enhanced channel throughput
for the same shear rate in the covering liquid. 
Fig.~\ref{fig:flux} also shows how, 
in the case that the covering liquid and the liquid in the channel have the
same viscosity, the efficiency of
the shear driving increases with increasing filling level.
This observation is in agreement with the corresponding one in body force 
driven systems
\cite{koplik06a}\/. Our MD simulations have encouragingly explored the
potential to miniaturize the concept of shear driving down to the
nano-scale.

Typical shear driven microfluidic devices might consist of two small
discs rotating relative to each other. One or both of the discs can
have a pattern of hydrophilic stripes on their hydrophobic surface. We
discus how with available technology such a 
device can be built with the size of a miniature hard disc, i.e.,
a few centimeters wide and long and a few millimeters thick.

As expected, for a contact angle larger than $90^\circ$ we observe
a surface tension driven pearling instability similar to the
Rayleigh-Plateau instability as discussed in detail for 
body force driven systems \cite{koplik06a}\/. In MD simulations of
a liquid completely wetting 
($\theta_{\text{eq}}=0^\circ$) a strip embedded into a non-wetting substrate
($\theta_{\text{eq}}=180^\circ$) the liquid
on the channel did not detach from the channel even if the shear was not
aligned with the channel direction. 
The issue of the shear rate at which the
liquid detaches from the channel is related to studies of moving droplets
with contact angle hysteresis \cite{dimitrakopoulos97,dimitrakopoulos98}\/. 

Our analysis suggests a significant application
potential 
for shear driven open microfluidic systems, accompanied by
further theoretical and experimental research.

\begin{acknowledgments}
We thank P. Dimitrakopoulos for fruitful discussions. M.R. acknowledges
financial support from the priority program SPP~1164 ``Micro and Nano
Fluidics'' of the Deutsche Forschungsgemeinschaft under grant
number RA~1061/2-1. J.K. is supported in part by the NASA
Exploration Systems Mission Directorate.  Computational resources
were provided by the NASA Advanced Supercomputing Division at the
Ames Research Center\/.
\end{acknowledgments}



\end{document}